\documentclass[aip,jmp,amsmath,amssymb,preprint]{revtex4-1}

\usepackage{natbib}

\begin{document}

\bibliographystyle{plainnat}

\title{Two-point Derivative Dispersion Relations}

\author{Erasmo Ferreira}
\email{erasmo@if.ufrj.br}
\affiliation{Instituto de F\'{\i}sica, Universidade Federal do Rio de Janeiro, 21941-972, Rio de Janeiro, Brasil}

\author{Javier Sesma}
\email{javier@unizar.es}
\affiliation{Departamento de F\'{\i}sica Te\'orica, Facultad de Ciencias, 50009, Zaragoza, Spain}

\date{\today}

\begin{abstract}
A new derivation is given for the representation, under certain conditions, of the integral dispersion relations of scattering theory through local forms. The resulting expressions have been obtained through an independent procedure to construct the real part, and consist of new mathematical structures of double infinite summations of derivatives. In this new form the derivatives are calculated at the generic value of the energy $E$ and separately at the reference point $E=m$ that is the lower limit of the integration. This new form may be more interesting in certain circumstances and directly shows the origin of the difficulties in convergence that were present in the old truncated forms called standard-DDR.
For all cases in which the reductions of the double to single sums were obtained in our previous work, leading to explicit demonstration of convergence, these new expressions are seen to be identical to the previous ones.
We present, as a glossary, the most simplified explicit results for the DDR's in the cases of imaginary amplitudes of forms $(E/m)^\lambda[\ln (E/m)]^n$, that cover the cases of practical interest in particle physics phenomenology at high energies. We explicitly study the expressions for the cases with $\lambda$ negative odd integers, that require identification of cancelation of singularities, and provide the corresponding final results.
\end{abstract}

\pacs{13.85.Dz; 13.85.Lg; 13.85.-t}

\keywords{hadronic scattering; dispersion relations; high energies}

\maketitle

\section{Introduction}

The displacement of the energy frontier of hadronic collisions in accelerators and cosmic ray experiments \cite{pdg} activates the importance of the studies of the soft diffractive and elastic processes that dominate the forward scattering amplitudes. The total cross section and the characteristic parameters of the forward elastic differential cross section define the energy dependence of the terms of real and imaginary amplitudes that are subject to the constraints determined by the general principles of causality, analyticity and crossing. Once the energy dependence of the forward ($t=0$) imaginary amplitude is given as input based on experiments, dispersion relations provide the corresponding terms of the real amplitude. Combined with the interference between the hadronic and Coulomb contributions, dispersion relations provide essential methods for the disentanglement of the observed data in terms of amplitudes. The knowledge obtained is needed for the control of theoretical models that must describe the observations in terms of dynamical mechanisms.

The integral dispersion relations (IDR) have the form of principal value integrations over infinite energy ranges. The manipulation of these mathematical instruments in the course of data analyses or of phenomenological studies, with many terms and parameters to be determined, is not straightforward. The derivative dispersion relations (DDR) convert integrations over infinite ranges into infinite series of derivatives at isolated energy points. The representations are given as double infinite series, whose convergence has been proved \cite{fs08} for the general case of interest of particle physics phenomenology at high energies, where the imaginary part of the amplitude is given as a sum
\begin{equation}
{\rm Im}\,F = \sum_{n} c_n\, (E/m)^{\lambda_n}\,[\ln (E/m)]^{k_n},  \qquad
k_n \; \mbox{nonnegative integer}. \label{dux1}
\end{equation}
For this general expression, the DDR obey the well known convergence criteria
 \cite{ed,fk1,kol1,kol2} and the expressions (27) and (28) of Ref. 2, already reduced
to simple convergent series, can be built.

Besides the usual dispersion relations connecting imaginary and real amplitudes, new relations have been introduced, obtained by derivation with respect to the momentum transfer $t$. \cite{e07} These dispersion relations for slopes (DDR/slopes) make
the connection between the slope of the real amplitude and the other forward scattering parameters. Once the imaginary slope $B_I$ is given with form $B_I(s)=C_0+C_I\,\ln s$ linear in $\ln s$, the forms to be calculated are also included in the general type of Eq. (\ref{dux1}) and are convergent. This possibility reinforces the importance of the use of the method of the derivative dispersion relations in the present studies of high energy scattering.

In our previous work \cite{fs08} we started from the expressions for the complete DDR derived by \'Avila and Menon \cite{am07} and obtained explicitly convergent forms for imaginary amplitudes of the form of Eq. (\ref{dux1}). We then presented numerical computations of all possible accuracy showing the exactness, in the cases considered, of the transformation
of IDR into DDR, and exhibited the reasons why the old truncated standard-DDR are inadequate, as they introduce singularities that do not exist in the original IDR.

In the present work we give a new derivation for the complete DDR, starting from first principles [namely from the original IDR] and also arrive at double infinite sums, different from those obtained before. \cite{am07} Besides being remarkably simpler, these new forms have the advantage of exhibiting clearly  the place where spurious singularities appeared in the truncated forms of standard-DDR.

Our derivation requires the imaginary part of the amplitude to fulfill some conditions that, obviously, should not be satisfied in the most general case, especially if one tries to take account properly of possible resonances in the low energy region. Nevertheless, the difficulties in analyzing the data, subject to errors, and the interest in reducing the number of adjustable parameters have led at high energies
to empirical amplitudes that basically reduce to entire functions of the logarithm of the energy.
We then concentrate on special cases included in Eq. (\ref{dux1}), for which
reduction to single infinite series are obtained. For these cases we obtain results that are mathematically identical to everything written before. \cite{fs08}

Using properties of summations involving Bernoulli numbers, we can put the results in more familiar forms, involving only trigonometric functions. This simplification that writes all representations of the real amplitudes in terms of elementary quantities
is presented in a compilation of formulas, given in Sect. V. This table of results is probably sufficient for the practical use in phenomenology of high energy scattering, including the DDR for slopes, where a $[\ln (E/m)]^3$ term appears. \cite{e07} In this list of results much attention is called to the required special treatment of cases with odd negative powers $\lambda$.

\section{The two-point Derivative Dispersion Relations}

We are interested in replacing the well known IDR (omitting subtraction constants)
\begin{equation}
{\rm Re}\,F_+(E,t) =
\frac{2E^2}{\pi}P\int_{m}^{+\infty}dE^\prime \, \frac{ {\rm Im}\, F_+(E^\prime,t)}{E^\prime(E^{\prime 2}-E^2)}  \label{inteven}
\end{equation}
and
\begin{equation}
{\rm Re}\,F_-(E,t)=
\frac{2E}{\pi}P\int_{m}^{+\infty}dE^\prime \, \frac{ {\rm Im}\, F_-(E^\prime,t)}{(E^{\prime 2}-E^2)} \, ,  \label{intodd}
\end{equation}
by equivalent DDR. Although it is usual to carry different expressions for the dispersion relations obeyed by the even and odd parts of the scattering amplitude, a formulation valid for the two cases can be used. In fact, both expressions Eqs. (\ref{inteven}) and (\ref{intodd}) can be written as
\begin{equation}
{\rm Re}\,G(E)=\frac{2E}{\pi}\,P\int_{m}^{+\infty}dE^\prime \,\frac{1}{E^{\prime 2}-E^2} \,{\rm Im}\,G(E^\prime), \label{tp3}
\end{equation}
where
\begin{equation}
G(E)= F_+(E,t)/E\qquad {\rm{or}} \qquad G(E)=F_-(E,t).  \label{destaque}
\end{equation}
Our goal is to replace the principal value integral in Eq. (\ref{tp3}) by a series of derivatives of ${\rm Im}\,G$. Previously obtained DDR, for instance the standard-DDR, deal with derivatives of ${\rm Im}\,G$ with respect to $\ln E$. It is, therefore, convenient to introduce the usual change of variable
\[
 \xi\equiv\ln(E/m), \qquad \xi^\prime\equiv\ln(E^\prime/m).
\]
Then, Eq. (\ref{tp3}) turns into
\begin{equation}
{\rm Re}\,G(E)=\frac{2}{\pi}\,P\int_{0}^{+\infty}d\xi^\prime
\,\frac{1}{e^{(\xi^\prime-\xi)}-e^{(\xi-\xi^\prime)}} \, {\rm Im}\,G(me^{\xi^\prime}). \label{tp4}
\end{equation}
The strategy commonly used to obtain DDR from the last equation runs along the following steps (see, for instance, Ref. 8): (i) an integration by parts, (ii) substitution of part of the integrand by its Taylor expansion (iii) integration, term by term, of the resulting expansion.
The second step is not free of controversy, as its validity imposes strict conditions to the function to be expanded. In the present work we avoid such kind of steps: the only expansion to be used is the geometric series, whose convergence conditions are well known.

According to the definition of principal value integral, the right hand side of Eq. (\ref{tp4}) is the limit of the sum of two integrals
\begin{equation}
{\rm Re}\,G(E)=\frac{2}{\pi}\,\lim_{\varepsilon\to 0^+}\,(I_1(\varepsilon)+I_2(\varepsilon)),   \label{tp5}
\end{equation}
where
\begin{equation}
I_1(\varepsilon)  = \int_{\xi+\varepsilon}^{+\infty}\!\! d\xi^\prime \,\frac{1}{e^{(\xi^\prime-\xi)}-e^{(\xi-\xi^\prime)}} \,{\rm Im}\,G(me^{\xi^\prime}),    \label{tp5a}
\end{equation}
\begin{equation}
I_2(\varepsilon)  = \int_{0}^{\xi-\varepsilon}\!\! d\xi^\prime \,\frac{1}{e^{(\xi^\prime-\xi)}-e^{(\xi-\xi^\prime)}} \,
{\rm Im}\,G(me^{\xi^\prime}),     \label{tp5b}
\end{equation}
that we are going to treat separately, although the procedure  to be used is the same in both cases: replace the integrand by its series expansion and integrate by parts each one of the terms of the series. However, convergence reasons require the use of a different expansion in each case. We find convenient to rewrite Eqs (\ref{tp5a}) and (\ref{tp5b}) in the forms
\begin{equation}
I_1(\varepsilon)  = \int_{\xi+\varepsilon}^{+\infty}d\xi^\prime \,\frac{e^{-(\xi^\prime-\xi)}}{1-e^{-2(\xi^\prime-\xi)}} \,{\rm Im}\,G(me^{\xi^\prime}),  
\quad
I_2(\varepsilon)  = \int_{0}^{\xi-\varepsilon}d\xi^\prime \,\frac{-e^{(\xi^\prime-\xi)}}{1-e^{2(\xi^\prime-\xi)}} \,
{\rm Im}\,G(me^{\xi^\prime}),     \label{tp7}
\end{equation}
and then the expansion known as geometric series can be used to obtain
\begin{equation}
I_1(\varepsilon)  = \int_{\xi+\varepsilon}^{+\infty}\!\! d\xi^\prime\sum_{p=0}^{\infty} e^{-(2p+1)(\xi^\prime-\xi)} \,
{\rm Im}\,G(me^{\xi^\prime}),   
\quad
I_2(\varepsilon)  = - \int_{0}^{\xi-\varepsilon}\!\! d\xi^\prime\sum_{p=0}^\infty  e^{(2p+1)(\xi^\prime-\xi)} \,
{\rm Im}\,G(me^{\xi^\prime}).     \label{tpp9}
\end{equation}
Now we require the imaginary part of the amplitude and its successive derivatives with respect to $\ln E$, that we represent by
\begin{equation}
{\rm Im}\,G^{(k)}(me^\xi) \equiv \frac{d^k }{d\xi^k}{\rm Im}\,G(me^\xi)\,,  \label{tpp10}
\end{equation}
to exist in the interval $0\leq\xi<+\infty$, that is, we assume
\begin{equation}
{\rm Im}\,G\in C^\infty  \left([m,+\infty)\right).   \label{tpp11}
\end{equation}
Besides this, we require all derivatives to behave at infinity in such a way that
\begin{equation}
\lim_{\xi\to +\infty} e^{-\xi}\,{\rm Im}\,G^{(k)}(me^\xi)\to 0\,, \qquad k=0, 1, 2, \ldots . \label{tpp12}
\end{equation}
Then, the uniformly convergent series in the right hand side of the two Eqs. (\ref{tpp9}) may be integrated term by term (see Ref. 9, $\S$ 4.7) to give
\begin{equation}
I_1(\varepsilon)  = \sum_{p=0}^{\infty}\int_{\xi+\varepsilon}^{+\infty}\!\! d\xi^\prime \,e^{-(2p+1)(\xi^\prime-\xi)} \,
{\rm Im}\,G(me^{\xi^\prime}),   
\quad
I_2(\varepsilon)  = - \sum_{p=0}^\infty\int_{0}^{\xi-\varepsilon}\!\! d\xi^\prime \,e^{(2p+1)(\xi^\prime-\xi)} \,
{\rm Im}\,G(me^{\xi^\prime}).     \label{tp9}
\end{equation}
Repeated integration by parts leads to
\begin{equation}
I_1(\varepsilon)  = \sum_{p=0}^\infty\sum_{k=0}^\infty \frac{-1}{(2p+1)^{k+1}}\,
\Bigg[e^{-(2p+1)(\xi^\prime-\xi)} \,{\rm Im}\,G^{(k)}(me^{\xi^\prime})\Bigg]_{\xi^\prime =\xi+\varepsilon}^{\xi^\prime\to +\infty},   \label{tp10}
\end{equation}
\begin{equation}
I_2(\varepsilon)  =  \sum_{p=0}^\infty\sum_{k=0}^\infty \frac{(-1)^{k+1}}{(2p+1)^{k+1}}\,
\Bigg[ e^{(2p+1)(\xi^\prime-\xi)} \,{\rm Im}\,G^{(k)}(me^{\xi^\prime})\Bigg]_{\xi^\prime =0}^{\xi^\prime=\xi-\varepsilon}.     \label{tp11}
\end{equation}
Substitution of these expressions in Eq. (\ref{tp5}) gives, in view of assumptions Eqs. (\ref{tpp11}) and (\ref{tpp12}),
\begin{equation}
{\rm Re}\,G(E)=\frac{2}{\pi}\sum_{p=0}^\infty\sum_{k=0}^\infty \left(\frac{2}{(2p+1)^{2k+2}}\,{\rm Im}\, G^{(2k+1)}(E)+\frac{(-1)^{k}}{(2p+1)^{k+1}}\,\left(\frac{E}{m}\right)^{-(2p+1)}{\rm Im}\,G^{(k)}(m)
\right).  \label{tp12}
\end{equation}
It is then immediate to write expressions for ${\rm Re}\,F_+(E,t)$ and ${\rm Re}\,F_-(E,t)$ in terms of derivatives of the corresponding ${\rm Im}\,G$ at the points $E$ and $m$.

A discussion of the convergence of the double series in Eq. (\ref{tp12}) does not seem to be easy for general Im $G(E)$. In the (truncated) standard-DDR, only the first term in the parenthesis is taken into account and Re $G(E)$ is approximated by
\begin{equation}
{\rm Re}\,G_1(E)=\frac{2}{\pi}\sum_{p=0}^\infty\sum_{k=0}^\infty \frac{2}{(2p+1)^{2k+2}}\,{\rm Im}\, G^{(2k+1)}(E),  \label{tpp13}
\end{equation}
that may be summed formally to give
\begin{equation}
{\rm Re}\,G_1(E)=\tan\left(\frac{\pi}{2}\frac{d}{d\ln E}\right){\rm Im}\,G(E).  \label{tpp14}
\end{equation}
The convergence of the tangent series was analyzed by Kol\'a\v{r} and Fischer. \cite{kol1} They have established in their Theorem 1 that the series converges for a given value of $E$ if and only if the series
\begin{equation}
\sum_{n=0}^\infty {\rm Im}\,G^{(2n+1)}(E)         \label{tpp15}
\end{equation}
is convergent. The theorem is correct, of course, but its application is limited to the truncated DDR, that in occasions is a bad approximation to the complete DDR. Let us consider, for instance, an amplitude such that
\begin{equation}
{\rm Im}\,G(E)=m/E. \label{tpp16}
\end{equation}
Obviously, the IDR is applicable and gives a finite value for Re $G(E)$, namely
\begin{equation}
{\rm Re}\,G(E)=-\frac{1}{\pi}\,\frac{m}{E}\,\ln\left(\frac{E^2}{m^2}-1\right). \label{tpp17}
\end{equation}
However, as
\begin{equation}
{\rm Im}\,G^{(2n+1)}(E)=-m/E, \label{tpp18}
\end{equation}
the series in Eq. (\ref{tpp15}) and, consequently, the tangent series become divergent. To remedy this, the above mentioned authors have proposed, in a mathematically rigorous paper, \cite{fk2} a kind of sum $\grave{{\rm a}}$ la Borel of the tangent series. The puzzle, however, was solved satisfactorily, in our opinion, by \'Avila and Menon \cite{am07}. Following a representation of Cudell, Martinov and Selyugin \cite{cude}, they found corrections due to a non-vanishing value of $m$,  that in our notation would read
\begin{eqnarray}
{\rm Re}\,G_2(E)) &=&
-\frac{1}{\pi}\ln\left|\frac{m-E}{m+E}\right|{\rm Im}\,G(m)   \nonumber \\
&+&\frac{2}{\pi} \sum_{k=0}^{\infty}\sum_{p=0}^{\infty}
\frac{(-1)^{k+1}\Gamma(k+1,(2p+1)\ln(E/m))}{(2p+1)^{k+2}\,k!}
\frac{d^{k+1}}{d (\ln E)^{k+1}} {\rm Im}\, G(E) \,,   \label{tpp19}
\end{eqnarray}
and which added to Re $G_1(E)$ would give the complete DDR. Here, following a different procedure, we have found, in Eq. (\ref{tp12}), an alternative expression for these corrections,
\begin{equation}
{\rm Re}\,G_2(E)=\frac{2}{\pi}\sum_{p=0}^\infty\sum_{k=0}^\infty \frac{(-1)^{k}}{(2p+1)^{k+1}}\,\left(\frac{E}{m}\right)^{-(2p+1)}{\rm Im}\,G^{(k)}(m)
\,,  \label{tpp20}
\end{equation}
as a double sum that could be reduced formally to a single sum
\begin{equation}
{\rm Re}\,G_2(E)=\frac{2}{\pi}\sum_{p=0}^\infty \left(\frac{E}{m}\right)^{-(2p+1)} \left[\frac{1}{2p+1+\frac{d}{d\ln E}} \,{\rm Im}\,G(E)\right]_{E=m}\,.  \label{tpp21}
\end{equation}
Calculations with particular forms of Im$\,G$, as those to be considered in the next sections, confirm the equivalence of the two different expressions, Eqs. (\ref{tpp19}) and (\ref{tpp21}), for  the corrections to the truncated DDR.

In analogy with Theorem 1 of Ref. 5, one could demonstrate that the double series in Eq. (\ref{tpp20}) is convergent if and only if the series
\begin{equation}
\sum_{k=0}^\infty (-1)^k\,{\rm Im}\,G^{(k)}(m)         \label{tpp22}
\end{equation}
is convergent. The proof would run along the same lines that the mentioned Theorem 1. The right hand side of Eq. (\ref{tpp20}) is of the form
\begin{equation}
\sum_{k=0}^\infty a_k\,(-1)^k\,{\rm Im}\,G^{(k)}(m)         \label{tpp23}
\end{equation}
with coefficients
\begin{eqnarray}
a_k & =\dfrac{2}{\pi} & \sum_{p=0}^\infty \frac{(m/E)^{2p+1}}{(2p+1)^{k+1}}   \nonumber \\
 & = & \frac{1}{\pi}\,\big({\rm Li}_{k+1}(m/E)-{\rm Li}_{k+1}(-m/E)\big),     \label{tpp24}
\end{eqnarray}
in terms of polylogarithms. It is then trivial to check that the sequence $\{a_k\}$ is bounded and decreasing for increasing $k$. Invocation of the Abel's test for convergence (see Ref. 9, $\S$ 2.31) would complete the proof. But we do not consider such a theorem to be of great help, as it would refer to the corrections, and not to the complete DDR. Coming back to the example in Eq. (\ref{tpp16}), the fact that
\begin{equation}
{\rm Im}\,G^{(k)}(m)=(-1)^k   \label{tpp25}
\end{equation}
makes the series in Eq. (\ref{tpp22}) to be divergent. Of course, Re $G_2(E)$ is in this case divergent, just to compensate the divergence found in Re $G_1(E)$ and to give a finite result for ${\rm Re}\,G(E)={\rm Re}\,G_1(E)+{\rm Re}\,G_2(E)$.

Summing up, we have obtained a two-point DDR, which adopts the explicit form Eq. (\ref{tp12}) or the more compact one
\begin{equation}
{\rm Re}\,G(E)=\tan\left(\frac{\pi}{2}\frac{d}{d\ln E}\right){\rm Im}\,G(E) + \frac{2}{\pi}\sum_{p=0}^\infty \left(\frac{E}{m}\right)^{-(2p+1)} \left[\frac{1}{2p+1+\frac{d}{d\ln E}} \,{\rm Im}\,G(E)\right]_{E=m}\,.  \label{tp14}
\end{equation}
For practical computations, Eq. (\ref{tp14}) seems to be preferable, because the first double series in the right hand side of Eq. (\ref{tp12}) converges very slowly. Nevertheless, Eq. (\ref{tp12}) is useful to show the cancelation of singularities in the first and second double series.

\section{A particular case: ${\rm Im}\,G(E) = (E/m)^\lambda$}

A particularly interesting possible form of ${\rm Im}\,G(E)$ is
\begin{equation}
{\rm Im}\,G(E) = (E/m)^\lambda. \label{tp15}
\end{equation}
Substitution of this form in Eq. (\ref{tp14}) gives
\begin{equation}
{\rm Re}\,G(E)=\left(\frac{E}{m}\right)^\lambda \tan\left(\frac{\pi}{2}\lambda\right) + \frac{2}{\pi}\sum_{p=0}^\infty \left(\frac{E}{m}\right)^{-(2p+1)}\!\!\frac{1}{2p+1+\lambda}\,.  \label{tp16}
\end{equation}
Both first and second terms in the right hand side of this equation are well defined for values $-1<\lambda<1$. The first term corresponds to the old standard-DDR, while the second one represents the corrections due to a finite value of $m$. As it should be, it coincides with the expression of the corrections obtained by \'Avila and Menon, \cite{am07} which were thoroughly discussed in our preceding paper. \cite{fs08} The first term becomes infinite for $\lambda=1$, as it is expected since the integral in the right hand side of Eq. (\ref{tp3}), with ${\rm Im}\,G(E^\prime)$ replaced by $(E^\prime/m)^\lambda$, becomes divergent for $\lambda=1$. One should not try to extend Eq. (\ref{tp16}) to values $\lambda>1$ because, in this case, Eq. (\ref{tp14}) does not follow from Eq. (\ref{tp12}). In fact, the first double sum in the right hand side of Eq. (\ref{tp12}) reads in this particular case
\[
\frac{2}{\pi}\,\sum_{p=0}^\infty\sum_{k=0}^\infty  \frac{2}{(2p+1)^{2k+2}}\left(\frac{E}{m}\right)^\lambda \lambda^{2k+1},
\]
which in view of the known relation \cite{prud}
\begin{equation}
\sum_{p=0}^\infty \frac{2}{(2p+1)^{2k+2}} = \frac{2^{2k+2}-1}{(2k+2)!}\,\pi^{2k+2}\,|B_{2k+2}|, \label{tp25}
\end{equation}
where the $B_{n}$ represent the Bernoulli numbers, can be written as
\[
\frac{2}{\pi}\left(\frac{E}{m}\right)^\lambda \sum_{k=0}^\infty \frac{2^{2k+2}-1}{(2k+2)!}\,|B_{2k+2}|\,\pi^{2k+2}\lambda^{2k+1}.
\]
The resulting series is divergent for $\lambda>1$ and cannot be written in terms of $\tan (\pi\lambda/2)$.

The situation is quite different for $\lambda =-1$, as the integral in Eq. (\ref{tp3}) is finite. The fact that both first and second terms in Eq. (\ref{tp16}) become divergent is related to the fact that such integral, which goes from $m$ to infinity, is represented by the difference of two integrals: one from 0 to infinity (the first term) and another one from 0 to $m$ (the opposite of second term). Both these integrals are divergent due to the behaviour of the integrand at $E=0$. This fact manifests itself as simple poles in each one of the two terms in the right hand side of Eq. (\ref{tp16}), considered as functions of $\lambda$. But the sum of the two terms is free of singularities. In practice, for $\lambda=-1$, one obtains from Eq. (\ref{tp16})
\begin{eqnarray}
{\rm Re}\,G(E)& = & \lim_{\lambda\to-1}\,\left[\left(\frac{E}{m}\right)^{\lambda}\,\tan\left(\frac{\pi}{2}\lambda\right)+ \frac{2}{\pi}\,\left(\frac{E}{m}\right)^{-1}\,\frac{1}{1+\lambda}\right] + \frac{2}{\pi}\sum_{p=1}^\infty \left(\frac{E}{m}\right)^{-(2p+1)}\frac{1}{2p}\nonumber  \\
 & = & -\,\frac{2}{\pi}\left(\frac{E}{m}\right)^{-1}\,\ln \left(\frac{E}{m}\right) - \frac{1}{\pi}\, \left(\frac{E}{m}\right)^{-1}\,\ln\left(1-\frac{m^2}{E^2}\right)  \nonumber \\
 & = & -\, \frac{1}{\pi}\,\frac{m}{E}\, \ln\left(\frac{E^2}{m^2}-1\right). \label{tp18}
\end{eqnarray}
Since the right hand side of Eq. (\ref{tp16}) turns out to be regular at $\lambda =-1$, that expression of ${\rm Re}\,G(E)$, although obtained for the interval $-1<\lambda<1$, can be continued to values of $\lambda$ to the left of $-1$ up to $\lambda=-3$, a value for which singularities appear again. It is not difficult to see that, analogously to what we have found for $\lambda=-1$, the singularities in the right hand side of Eq. (\ref{tp16}) occurring for $\lambda=-(2N+1)$, $N=1, 2, \ldots$, balance to a finite value. The conclusion is that it can be used for odd negative values of $\lambda$  with replacement of Eq. (\ref{tp16}) by
\begin{eqnarray}
{\rm Re}\,G(E) &=& \frac{2}{\pi}\left(-\left(\frac{E}{m}\right)^{-(2N+1)}\ln \left(\frac{E}{m}\right) + {\sum_{p=0,\, p\neq N}^\infty} \left(\frac{E}{m}\right)^{-(2p+1)}\frac{1}{2p-2N}\right),  \nonumber  \\
& & \hspace{3cm} \mbox{for} \quad  \lambda=-(2N+1), \quad N=0,1, 2, \ldots.  \label{tp19}
\end{eqnarray}

\section{A more general case}

Let us consider now the case of ${\rm Im}\,G(E)$ being a combination of terms of the form $(E/m)^\lambda\left[\ln(E/m)\right]^n$, namely
\begin{equation}
{\rm Im}\,G(E)=\sum_i c_i\,(E/m)^{\lambda_i}\left[\ln(E/m)\right]^{n_i}, \qquad n_i\quad\mbox{nonnegative integer}.  \label{tp20}
\end{equation}
The linearity of dispersion relations allows one to tackle the problem by studying the DDR, separately, for each one of the terms in the right hand side of Eq. (\ref{tp20}). Therefore, let us consider the case
\begin{equation}
{\rm Im}\,G(E)=(E/m)^\lambda\left[\ln(E/m)\right]^n, \qquad n\quad\mbox{positive integer}.  \label{tp21}
\end{equation}
Since
\begin{equation}
(E/m)^\lambda\left[\ln(E/m)\right]^n = \frac{\partial^n}{\partial \lambda^n}(E/m)^\lambda,  \label{tp22}
\end{equation}
the DDR for the case at hand stems from Eq. (\ref{tp16}) by repeated derivation, $n$ times, with respect to $\lambda$, to give
\begin{equation}
{\rm Re}\,G(E)=\left(\frac{E}{m}\right)^\lambda \sum_{j=0}^n {n \choose j} \left[\ln \left(\frac{E}{m}\right)\right]^{n-j} \frac{\partial^j}{\partial \lambda^j}\tan\left(\frac{\pi}{2}\lambda\right) + \frac{2}{\pi}\sum_{p=0}^\infty \left(\frac{E}{m}\right)^{-(2p+1)}\!\!\!\! \frac{(-1)^n \,n!}{(2p+1+\lambda)^{n+1}}\, .  \label{tp23}
\end{equation}
This expression is well defined for $-1<\lambda<1$. It cannot be extended to values $\lambda\geq 1$, where the integral in Eq. (\ref{tp3}) becomes divergent. It can, instead, be continued to negative values of $\lambda$ in spite of the singularities (of order $n+1$) in both terms of the right hand side occurring for $\lambda=-(2N+1)$, $N=0, 1, 2, \ldots$. Analogously to what happens with Eq. (\ref{tp16}), the sum of singularities produces a finite result that can be obtained by a limit procedure.

In the particular case of $\lambda=0$, i. e., ${\rm Im}\,G(E)= [\ln (E/m)]^n$, the corresponding DDR can be trivially obtained from Eq. (\ref{tp23}), by taking the limit for $\lambda \to 0$, or, equivalently, by direct application of Eq. (\ref{tp12}). We obtain in this case
\begin{eqnarray}
{\rm Re}\,G(E)& = & \frac{2}{\pi}\sum_{p=0}^\infty\Bigg(\sum_{k=0}^{[(n-1)/2]} \frac{2}{(2p+1)^{2k+2}}\,\frac{n!}{(n-2k-1)!}\left(\ln\left(\frac{E}{m}\right)\right)^{n-2k-1} \nonumber  \\ & & \hspace{3cm}\  + \ \frac{(-1)^{n}\,n!}{(2p+1)^{n+1}}\,\left(\frac{E}{m}\right)^{-(2p+1)}\Bigg),  \label{tp24}
\end{eqnarray}
where the symbol [ ] in the upper limit of the summation in $k$ stands for integer part. By using again the relation in Eq. (\ref{tp25}),
we get finally, in terms of the Bernoulli numbers $B_n$,
\begin{equation}
{\rm Re}\,G(E) =  \frac{2\,n!}{\pi}\Bigg(\sum_{k=0}^{[(n-1)/2]} \frac{(2^{2k+2}-1)\,\pi^{2k+2}\,|B_{2k+2}|}{(2k+2)!(n-2k-1)!}\left[\ln\left(\frac{E}{m}\right)\right]^{n-2k-1} \!\!+ (-1)^{n}\sum_{p=0}^\infty\frac{(E/m)^{-(2p+1)}}{(2p+1)^{n+1}}\Bigg).  \label{tp26}
\end{equation}

\section{A compilation}

For practical purposes, we give below a compilation of the resulting DDR for the cases of ${\rm Im}\,G(E)$ given by ${\rm Im}\, G(E)=(E/m)^\lambda\left[\ln(E/m)\right]^n$, with four particular values of $n$, namely $n=0, 1, 2, 3$, and assuming $\lambda<1$ and $\lambda\neq-(2N+1)$, $N=0, 1, 2, \ldots $. Along the following equations, the abbreviation
\[
\alpha \equiv \pi\lambda/2
\]
is used. Remember that, as assumed in Eq. (\ref{destaque}), $G(E)$ represents either $F_+(E,t)/E$ or $F_-(E,t)$. We then have
\begin{eqnarray}
{\rm for}\quad n=0,\quad {\rm Re}\,G(E)& = & \left(\frac{E}{m}\right)^\lambda \tan\alpha + \frac{2}{\pi}\sum_{p=0}^\infty
\frac{(E/m)^{-(2p+1)}}{2p+1+\lambda}\, ,  \label{tp27}  \\
{\rm for}\quad n=1,\quad {\rm Re}\,G(E) & = & \left(\frac{E}{m}\right)^\lambda\left[\ln\left(\frac{E}{m}\right)\tan\alpha + \frac{\pi}{2}\sec^2\alpha\right]
- \frac{2}{\pi}\sum_{p=0}^\infty\frac{(E/m)^{-(2p+1)}}{(2p+1+\lambda)^2}\,, \label{tp28} \\
{\rm for}\quad n=2,\quad {\rm Re}\,G(E) & = & \left(\frac{E}{m}\right)^\lambda\left[\left[\ln\left(\frac{E}{m}\right)\right]^2 \tan\alpha
+ \pi\sec^2\alpha\left(\ln\left(\frac{E}{m}\right)+\frac{\pi}{2}\tan\alpha\right)\right] \nonumber \\
& & \hspace{4cm}\ + \ \frac{4}{\pi}\sum_{p=0}^\infty\frac{(E/m)^{-(2p+1)}}{(2p+1+\lambda)^3}\,, \label{tp29} \\
{\rm for}\quad n=3,\quad {\rm Re}\,G(E) & = & \left(\frac{E}{m}\right)^\lambda \Bigg[\left[\ln\left(\frac{E}{m}\right)\right]^3 \tan\alpha   \nonumber\\
& & \hspace{-2cm}\ + \ \frac{\pi}{2}\sec^2\alpha\left(3\left[\ln\left(\frac{E}{m}\right)\right]^2
+ 3\pi\ln\left(\frac{E}{m}\right)\tan\alpha+\frac{\pi^2}{2}\left(1+3\tan^2\alpha\right)\right)\Bigg]  \nonumber \\
& & \hspace{4cm} \ - \ \frac{12}{\pi}\sum_{p=0}^\infty\frac{(E/m)^{-(2p+1)}}{(2p+1+\lambda)^4}\,. \label{tp30}
\end{eqnarray}
For electronic computational purposes, a function $\mathcal{S}(n,\lambda,E/m)$ can be programmed to represent the last term in the right hand side of each one of these equations. It would be written
\begin{equation}
\mathcal{S}(n,\lambda,E/m) = (-1)^n\,\frac{2\,n!}{\pi}\sum_{p=0}^\infty\frac{(E/m)^{-(2p+1)}}{(2p+1+\lambda)^{n+1}}\,. \label{extra1}
\end{equation}

In the case of $\lambda$ being a negative odd integer, $\lambda=-(2N+1)$, $N=0, 1, 2, \ldots $, the limit procedure must be followed, and we obtain
\begin{eqnarray}
{\rm for}\quad n=0,\quad {\rm Re}\,G(E) & = &- \frac{2}{\pi}\ln \left(\frac{E}{m}\right)\left(\frac{E}{m}\right)^{-(2N+1)} +  \frac{1}{\pi}{\sum_{p=0,\, p\neq N}^\infty} \frac{(E/m)^{-(2p+1)}}{p-N}\,, \label{tp31}  \\
{\rm for}\quad n=1,\quad {\rm Re}\,G(E) & = & \left(\frac{\pi}{6}-\frac{1}{\pi}\left[\ln\left(\frac{E}{m}\right)\right]^2\right)\left(\frac{E}{m}\right)^{-(2N+1)} \nonumber \\
& & \hspace{4cm}-\  \frac{1}{\pi}\,{\sum_{p=0,\,p\neq N}^\infty}\frac{(E/m)^{-(2p+1)}}{2\,(p-N)^2}\,, \label{tp32}  \\
{\rm for}\quad n=2,\quad {\rm Re}\,G(E) & = & \frac{1}{3}\ln\left(\frac{E}{m}\right)\left(\pi-\frac{2}{\pi}\left[\ln\left(\frac{E}{m}\right)\right]^2\right)\left(\frac{E}{m}\right)^{-(2N+1)}
\nonumber  \\
& & \hspace{4cm}\ +\  \frac{1}{\pi}\,{\sum_{p=0,\,p\neq N}^\infty}\frac{(E/m)^{-(2p+1)}}{2^2\,(p-N)^3}\,, \label{tp33}  \\
{\rm for}\quad n=3,\quad {\rm Re}\,G(E) & = & \left(\frac{\pi^3}{60}+\frac{\pi}{2}\left[\ln\left(\frac{E}{m}\right)\right]^2-\frac{1}{2\pi}\left[\ln\left(\frac{E}{m}\right)\right]^4\right)
\left(\frac{E}{m}\right)^{-(2N+1)}  \nonumber  \\
& & \hspace{4cm}\ -\  \frac{1}{\pi}\,{\sum_{p=0,\,p\neq N}^\infty}\frac{(E/m)^{-(2p+1)}}{2^3\,(p-N)^4}\,. \label{tp34}
\end{eqnarray}
The last term in the right hand side of each of these equations can be written in the general form
\begin{equation}
\mathcal{T}(n, N, E/m) = (-1)^n\,\frac{n!}{2^n\,\pi}\,{\sum_{p=0,\,p\neq N}^\infty}\frac{(E/m)^{-(2p+1)}}{(p-N)^{n+1}}\,,  \label{tp35}
\end{equation}
or, in terms of the polylogarithm function Li$_r$, also as
\begin{equation}
\mathcal{T}(n, N, E/m) =
\frac{n!}{2^n\,\pi}\,(E/m)^{-(2N+1)}\left(-\sum_{q=1}^{N}\frac{(E/m)^{2q}}{q^{n+1}}+(-1)^n\,{\rm Li}_{n+1}\left((m/E)^2\right)\right).  \label{tp36}
\end{equation}

\section{Conclusions}

The general treatment of the passage from the even and odd IDR of scattering theory to DDR, made without any explicit approximation or truncation, \cite{am07} has lead to expressions with double infinite series. The convergence of these complicated structures has to be proved in each case where it is to be used, and this has been cause of concern for some time. The expressions were successfully employed in the analysis of pp and p$\bar{{\rm p}}$ scattering at high energies, where specific forms of functions Im$\,F_+(E/m)$ and Im$\,F_-(E/m)$ appear, all included in the the form given in Eq. (\ref{dux1}), with practical demonstration of the identity of results using the original IDR and the DDR with double series.

Then, through the use of properties of series of hypergeometric functions,
it was proved \cite{fs08},
for all cases of functions included in Eq. (\ref{dux1}),
that the double sums can be reduced to single sums,  and it was verified that the resulting simplified expressions are indeed convergent. Besides, it was shown, with plenty of direct numerical examples, that the final representations obtained for the IDR in terms of local forms are exact, thus confirming the validity of the method leading to the double sums. The derivation of the exact expressions of DDR also exhibited the origin of the spurious singularities presented by the truncated forms, called standard-DDR, used up to that moment in applications in the scattering problem.

In the present work we have derived a new expression, also in the form of double infinite sums, to represent the dispersion integrals. The advantage of the new expression becomes evident form a comparison of our relatively simple Eq. (\ref{tpp20}) (or Eq. (\ref{tpp21})) with Eq. (\ref{tpp19}), which is rather cumbersome. The derivation of the new expression is made with usual and safe analytical methods, involving expansions, and the control of convergence requires the specification of the form of the function in the dispersion relation. The essential convergence property is proved in the present work, for the general form of Eq. (\ref{dux1}), with reduction from double to single series through elementary identities (no need of knowledge of sophisticated properties of transcendental functions). Of course, it is important to remark that the final expressions are identically the same as those obtained before.

As a formal development, very important for the practical use, we have now written the sums involving Bernoulli numbers in terms of trigonometric functions, using well known identities. With these elementary expressions, we have presented in Sec. V a compilation of results applicable to all forms appearing in the description of hadronic scattering at high energies.

Finally, we point out the importance  of the use of dispersion relations, in particular the very handy DDR, in the present studies of pp scattering at very high energies, where the parametrizations of the forward imaginary amplitude and its slope fall into simple dependences on $E/m$ and $\ln (E/m)$. Exact expressions for the DDR are essential tool for the disentanglement of the real and imaginary parts that enter the observable quantities.

\begin{acknowledgments}
Financial support from Conselho Nacional de Desenvolvimento
Cient\'{\i}fico e Tecnol\'{o}gico  (CNPq, Brazil) and from Departamento de Ciencia, Tecnolog\'{\i}a y Universidad del Gobierno de Arag\'on (Project E24/1) and Ministerio de Ciencia e Innovaci\'on (Project MTM2009-11154) is gratefully acknowledged.
\end{acknowledgments}

\end{document}